\pdfoutput=1
\documentclass[twocolumn,times]{aastex63}
\usepackage{graphicx}
\usepackage{mathtools,upgreek} 
\usepackage{natbib}
\usepackage{outlines}
\usepackage{amsmath}
\usepackage{verbatim}
\usepackage{appendix}
\usepackage{multirow,tabularx,array}
\usepackage{comment}
\usepackage[version=4]{mhchem} 

\newcolumntype{C}[1]{>{\centering\arraybackslash}m{#1}}

\submitjournal{ApJL}
\received{November 30, 2020}
\revised{December 17, 2020}
\accepted{December 21, 2020}

\newif\iftwocol
\twocolfalse

\begin{document}

\title{Complications in the ALMA Detection of Phosphine at Venus}

\author[0000-0001-8379-1909]{Alex B. Akins}
\affiliation{Jet Propulsion Laboratory, California Institute of Technology, Instruments Division, Pasadena, California 91011, USA}
\correspondingauthor{Alex Akins}
\email{alexander.akins@jpl.nasa.gov}

\author[0000-0003-0429-9487]{Andrew P. Lincowski}
\affiliation{Department of Astronomy and Astrobiology Program, University of Washington, Box 351580, Seattle, Washington 98195, USA}
\affiliation{NASA Nexus for Exoplanet System Science, Virtual Planetary Laboratory Team, Box 351580, University of Washington, Seattle, Washington 98195, USA}

\author[0000-0002-1386-1710]{Victoria S. Meadows}
\affiliation{Department of Astronomy and Astrobiology Program, University of Washington, Box 351580, Seattle, Washington 98195, USA}
\affiliation{NASA Nexus for Exoplanet System Science, Virtual Planetary Laboratory Team, Box 351580, University of Washington, Seattle, Washington 98195, USA}

\author[0000-0003-3962-8957]{Paul G. Steffes}
\affiliation{School of Electrical and Computer Engineering, Georgia Institute of Technology, Atlanta, GA, 30308, Atlanta, GA}

\begin{abstract}

Recently published ALMA observations suggest the presence of 20 ppb PH$_3$ in the upper clouds of Venus. This is an unexpected result, as PH$_3$ does not have a readily apparent source and should be rapidly photochemically destroyed according to our current understanding of Venus atmospheric chemistry. While the reported PH$_3$ spectral line at 266.94 GHz is nearly co-located with an SO$_2$ spectral line, the non-detection of stronger SO$_2$ lines in the wideband ALMA data is used to rule out SO$_2$ as the origin of the feature. We present a reassessment of wideband and narrowband datasets derived from these ALMA observations. The ALMA observations are re-reduced following both the initial and revised calibration procedures discussed by the authors of the original study. We also investigate the phenomenon of apparent spectral line dilution over varying spatial scales resulting from the ALMA antenna configuration. A 266.94 GHz spectral feature is apparent in the narrowband data using the initial calibration procedures, but this same feature can not be identified following calibration revisions. The feature is also not reproduced in the wideband data. While the SO$_2$ spectral line is not observed at 257.54 GHz in the ALMA wideband data, our dilution simulations suggest that SO$_2$ abundances greater than the previously suggested 10 ppb limit would also not be detected by ALMA. Additional millimeter, sub-millimeter, and infrared observations of Venus should be undertaken to further investigate the possibility of PH$_3$ in the Venus atmosphere.
\end{abstract}

\section{Introduction}
A recent paper by \citet{Greaves2020} reports the identification of PH$_3$ in the atmosphere of Venus via radio telescope observations of the PH$_3$ 1-0 transition line near 266.94 GHz (1.123 mm). A spectral feature was first observed near this frequency using the James Clerk Maxwell Telescope (JCMT), a single dish millimeter/sub-millimeter radio telescope operating in Mauna Kea, HI\footnote{While JCMT continues to operate, the specific receiver used for these observations has since been retired}. This spectral feature, however, could not be uniquely identified as PH$_3$ due to the presence of the SO$_2$ 30$_{9, 21}$-31$_{8, 24}$ line within 1.5 km/s of the PH$_3$ 1-0 line. Since SO$_2$ is an abundant trace gas in the Venus atmosphere at all altitudes (see \citet{Marcq2018} and references therein), it could be the case that the detected line was that of SO$_2$ instead of PH$_3$. To assess this uncertainty, the JCMT observations were followed up by interferometric observations of Venus near 266.94 GHz using the Atacama Large Millimeter Array (ALMA). The ALMA Band 6 correlator configurations for these observations included wideband (1.875 GHz bandwidth, 976.562 kHz spectral resoluton) and narrowband (117.1875 MHz bandwidth, 61.035 kHz spectral resolution) setups centered on both the PH$_3$ 1-0 line of interest and the nearby HDO 2$_{2,0}$-3$_{1,3}$ line for validation. The wideband configuration centered on the 266.94 GHz spectral feature can also be used to estimate the intensities of two stronger SO$_2$ tracer lines (13$_{3,11}$-13$_{2,12}$ near 267.54 GHz and 28$_{4,24}$-28$_{3,25}$ near 267.72 GHz). These tracer lines could then be used by the authors to determine if the spectral feature at 266.94 GHz could be attributed to PH$_3$. \par
Following data reduction and imaging, \citet{Greaves2020} report a non-detection of the SO$_2$ 13$_{3,11}$-13$_{2,12}$ tracer line, and they suggest an upper concentration limit of 10 ppb SO$_2$ based on the amplitude of the spectral ripple observed in the wideband data. Assuming this SO$_2$ abundance, the authors determine that the contamination from the weaker SO$_2$ 30$_{9, 21}$-31$_{8, 24}$ line is not sufficient to explain the observed spectral feature, and therefore, that the 266.94 line can be attributed to 20 ppb PH$_3$ (see Figure 4 in \citet{Greaves2020}). They also report detection of the 266.94 GHz spectral feature in both the wideband and the narrowband ALMA data, as well as identifying the expected HDO line in their other spectral configurations. \par 
Due to the weighting function of 266.94 GHz continuum emission peaking in the middle cloud layer of Venus (53-61 km above the mean surface), \citet{Greaves2020} propose that 20 ppb PH$_3$ is present in the clouds of Venus. Given the chemistry of the Venus atmosphere and the PH$_3$ formation and destruction rates from the simulations of \citet{Bains2020}), it is not expected that such a high concentration of PH$_3$ can exist with long-term stability without a significant, unaccounted for source (see Figure 5 in \citet{Greaves2020}. After ruling out several formation mechanisms based on the current understanding of Venus atmospheric chemistry, \citet{Greaves2020} suggest that the origin of PH$_3$ could be biogenic. A recent paper by \citet{Limaye2018} reviews hypotheses for biological activity within the Venus cloud deck in the context of observed UV absorption, and another recent paper by \citet{Seager2020} proposes a life cycle for such life through cycling of H$_2$SO$_4$ cloud aerosols. The identification of PH$_3$ in the cloud region, therefore, could be in favor of such a hypothesis. However, several recent papers call into question the detection of a spectral feature at 266.94 GHz \citep{Snellen2020, Villanueva2020, Thompson2020} and its interpretation as being due to PH$_3$ \citep{Encrenaz2020, Villanueva2020, Lincowski2020}.
The authors of the original study have also performed a reanalysis of their ALMA data following the identification of errors in their calibration procedures \citet{Greaves2020a}. Their reanalysis again reports detection of PH$_3$ with the ALMA data, but they report a lower abundance of 1 ppb. \citet{Greaves2020a} also identify image regions with apparently stronger absorption compared with the full disk-average.

\par

Given the implications of the identification of PH$_3$ as a trace gas in the Venus mesosphere, the observations reported by \citet{Greaves2020} must be carefully considered and independently verified by other research groups. In this paper, we discuss our independent analysis of the \citet{Greaves2020} ALMA observations which is complementary to the recent reanalyses of \citet{Snellen2020}, \citet{Villanueva2020}, and \citet{Greaves2020a}. We re-reduce and image the ALMA data using both the initial calibration methods of \citet{Greaves2020} and the revised calibration methods presented in their follow-on letter \citep{Greaves2020a}. We also perform model simulations to assess how different spatial extents of absorbing gases affect the apparent spectral line dilution (suppression of an absorption feature) that results from the observing geometry and configuration of the ALMA antennas.  While we are able to observe a 266.94 GHz feature in the ALMA narrowband data using the initial calibration procedures, we fail to detect it following the revised calibration. No 266.94 GHz spectral feature is observed in the wideband data using either set of calibration procedures. We determine the line dilution incurred by the ALMA observation geometry admits SO$_2$ concentrations that are considerably higher than the 10 ppb limit suggested by \citet{Greaves2020} if the gas is globally or hemispherically distributed. While we are unable to reproduce the detection of PH$_3$ in the ALMA data, we encourage new observations of Venus in the vicinity of PH$_3$ spectral lines to to further investigate the possibility of PH$_3$ at Venus.

\section{Methods}
ALMA observed Venus in the C-2 configuration (15-314 m baselines) on March 5, 2019 over two observing sessions, and the observation data were obtained from the ALMA archive. The Common Astronomy Software Applications (CASA) package was used for all data reduction and imaging of the ALMA data, as well as for simulations described in this section. The spectra of the resulting images of Venus were then analyzed using the Cube Analysis and Rendering Tool for Astronomy (CARTA). 

\subsection{Reduction and Imaging of ALMA Data}
The ALMA wideband and narrowband data centered on 266.94 GHz were initially reduced and imaged using the scripts provided by \citet{Greaves2020} (Supplemental Software 2 and 3). We did not investigate the wideband and narrowband datasets centered on the HDO line. System temperature, antenna position, and water vapor radiometer calibration tables were applied to the data prior to initial flagging. In the wideband spectrum, there is telluric contamination apparent at 267.3 GHz, but this contamination does not overlap with the other spectral lines of interest. During the data flagging stage, the datasets were duplicated. For one of these copies, baselines shorter than 33 m were removed from the data (this dataset will be referred to as LB for 'long baselines'), whereas all baselines were retained for the other copy (this dataset will be referred to as AB for 'all baselines'). The Venus observations are flux-calibrated and bandpass-calibrated using observations of Callisto prior to antenna gain/phase calibration using observations of J2000-1748. Following these calibrations, the data from the two observing sessions were combined to form the full narrowband and wideband datasets centered at 299.64 GHz. \par
Instead of using the 'Butler-JPL-Horizons 2012' model of Venus native to CASA as the starting model for self-calibration, we adopt a functionally similar approach by fitting a limb darkened disk model to the calibrated visibilities. This starting model is used for phase-only self-calibration with a solution interval of 30s (for the visibility model, see the Appendix A of \citet{Butler1999}). Here the ALMA datasets are again duplicated, and a first-order continuum is subtracted from one of the copies. As in Supplemental Software 3 from \citet{Greaves2020}, both the continuum-subtracted and continuum-included observations are imaged using a multiscale CLEAN approach. The resulting products are spectral image cubes of Venus with 1920 channels for both the wideband and narrowband data. We obtain disk-averaged spectra as well as spectra averaged over equators, mid-latitudes, and polar regions. These data are then spectrally averaged to a common resolution of 1.1 km/s, and the background spectral ripples are subtracted using polynomial functions. The line-to-continuum ratios are determined using the fluxes of the continuum images for both baseline configurations. For the narrowband data, 12th order polynomials are fit to the spectral baseline using a $|$40$|$ km/s window excluding the central $|$5$|$ km/s from the fit in the manner described in \citet{Greaves2020}. While we attempt to follow the methods of \citet{Greaves2020}, we note that the use of a high order polynomial to fit the narrowband data artificially reduces spectral noise in this dataset \citep{Snellen2020, Villanueva2020}.
\par 
Following the initial report by \citet{Greaves2020}, it was determined that certain procedures during the initial calibration with CASA affected the efficacy of the bandpass calibration and produced anomalously high spectral ripples in the resulting Venus images \citep{Villanueva2020, Greaves2020a}. The initial calibration procedures were then revised by the Joint ALMA Observatory, and these scripts, as well as revised image products, were made available on the ALMA archive. The most significant change was the use of a polynomial fit (3rd order amplitude, 5th order phase) to determine bandpass calibrations from the Callisto observations. We used these scripts to re-reduce the dataset, only modifying the archive procedures to remove antenna baselines less than 33 m following \citet{Greaves2020a}. Following these procedures, we produced new images of Venus with the specifications mentioned earlier.

\subsection{Simulation of ALMA Line Dilution}
The selection and configuration of the ALMA antennas during the observations places limits on the reliability of the reproduced large scale structure of Venus. The maximum recoverable spatial scale (MRS) of ALMA Band 6 observations using the C-2 antenna configuration is near 9.8 arcseconds \citep{Remijan2019}, which is smaller than the angular diameter of Venus at the time of observation (15.36 arcseconds). Furthermore, to reduce spectral ripples observed in their data, \citet{Greaves2020} remove data from baselines with shorter spacing than 33 meters (roughly 30k$\lambda$), reducing the MRS to 7.1 arcseconds. This complicates the interpretation of brightness temperature structure at larger angular scales (greater than mesoscale features in Venus’ atmosphere) and will most significantly impact spectral detection of trace gases that are distributed globally at Venus. \citet{Greaves2020} assessed the problem of line dilution in their data by adding a cloud of PH$_3$ with a diameter of 8 arcseconds to a model of the Venus disk. Imaging this model, \citet{Greaves2020} determined a line dilution of 80-90\%. This justified their decision to infer PH$_3$ abundance from the single disk JCMT observation, which does not suffer from line dilutions and would return a more accurate abundance. \par
This line dilution problem is important for SO$_2$, which is distributed globally at Venus and generally exhibits hemispheric variations in abundance \citep{Encrenaz2019}.
To explore spatial scale-dependent line dilution, the simulated SO$_2$ spectrum of Venus presented in Figure 3D of \citet{Lincowski2020} is imposed over several scales on a limb-darkened disk model of Venus. Since the ALMA configuration of \citet{Greaves2020} is more sensitive to flux from the limb, the spatial regions of SO$_2$ enhancement are distributed from the center of the disk to the limb. The Fourier Transform of these models are then re-imaged with CASA using the corresponding full-baseline coverage and the short-baseline exclusion of the \citet{Greaves2020} ALMA observations.

\section{Results}

\subsection{ALMA Data}
\begin{figure*}[tbh!]
    \centering
    \includegraphics[width=0.45\textwidth]{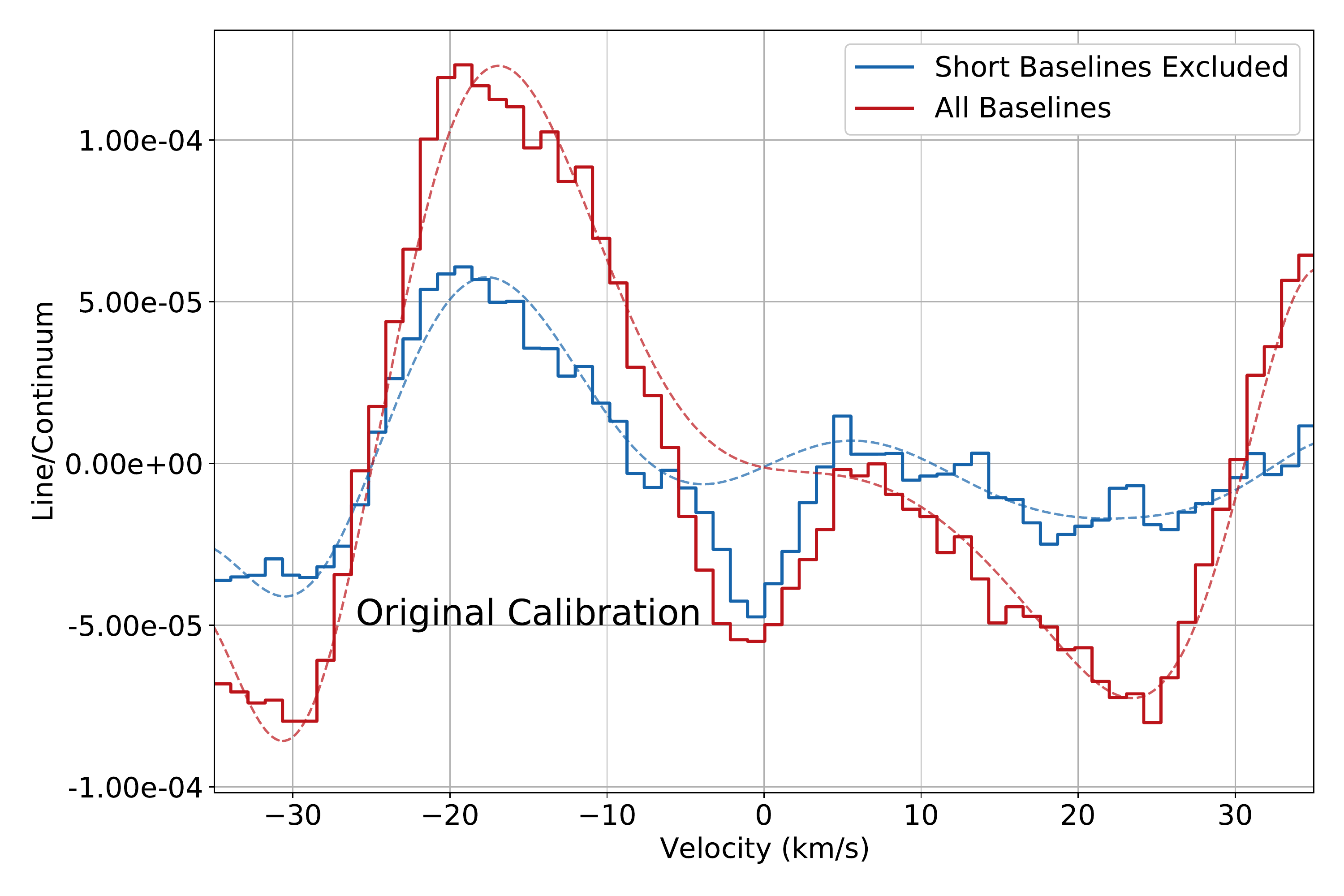}
    \includegraphics[width=0.45\textwidth]{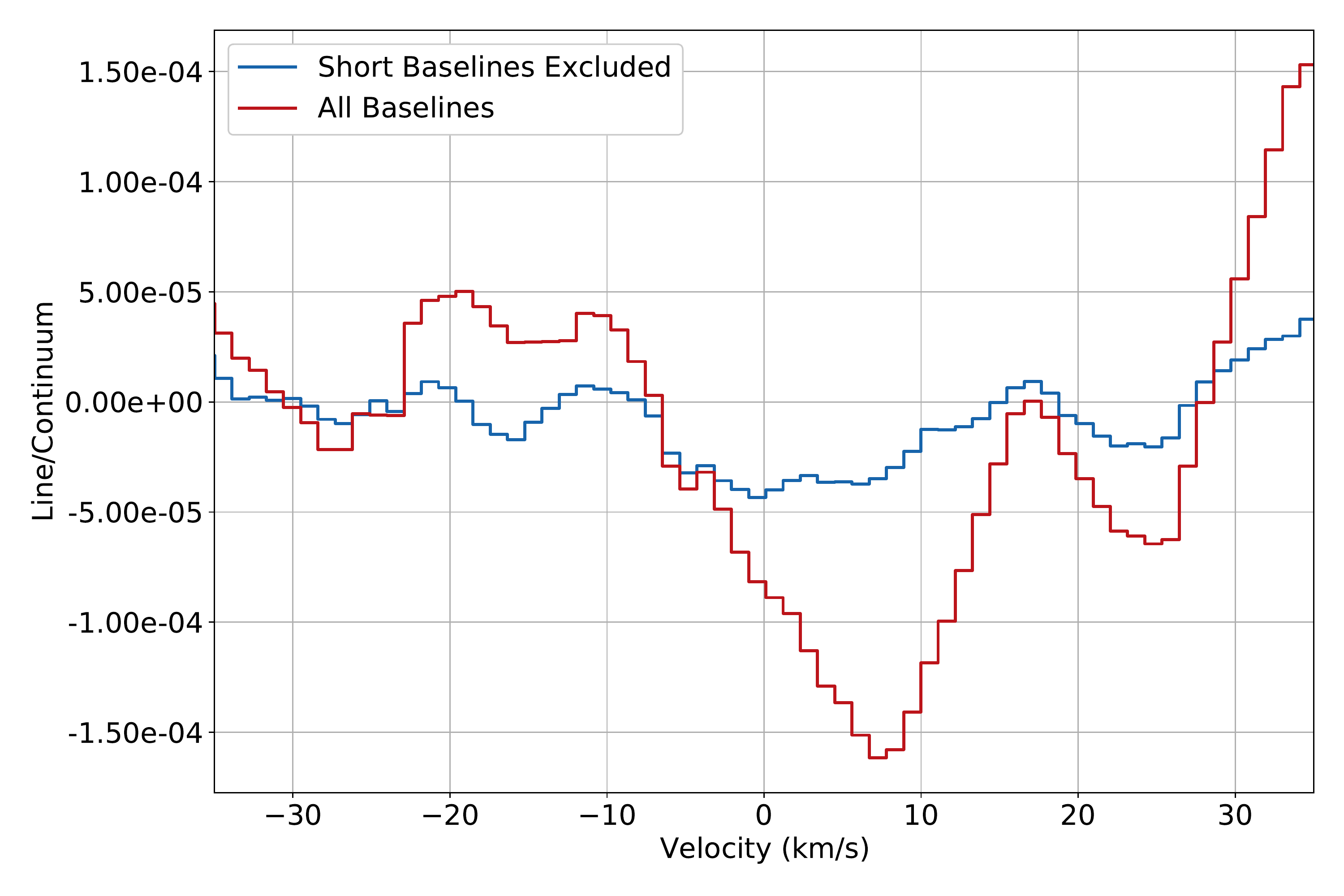} \hfill
    \includegraphics[width=0.45\textwidth]{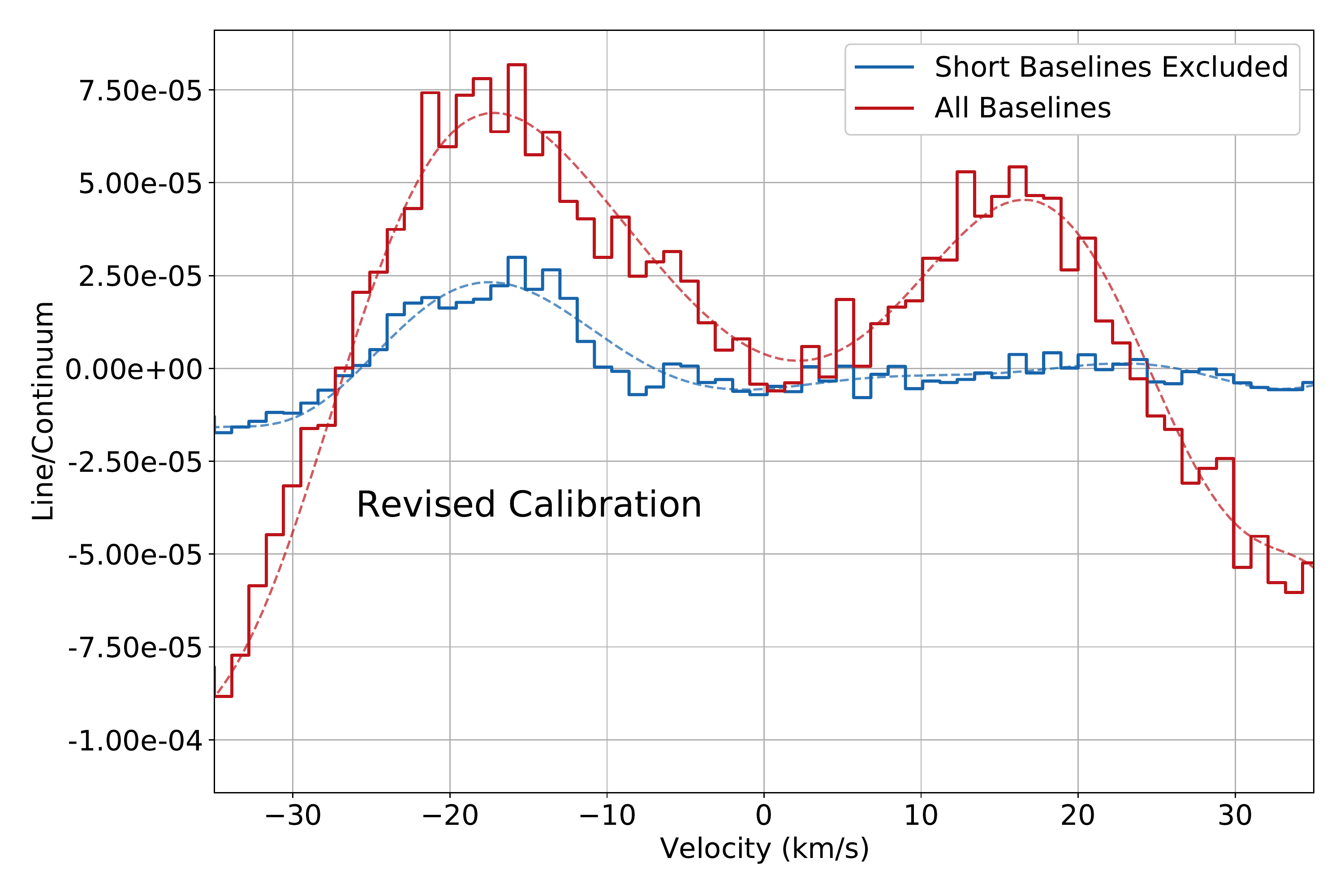}
    \includegraphics[width=0.45\textwidth]{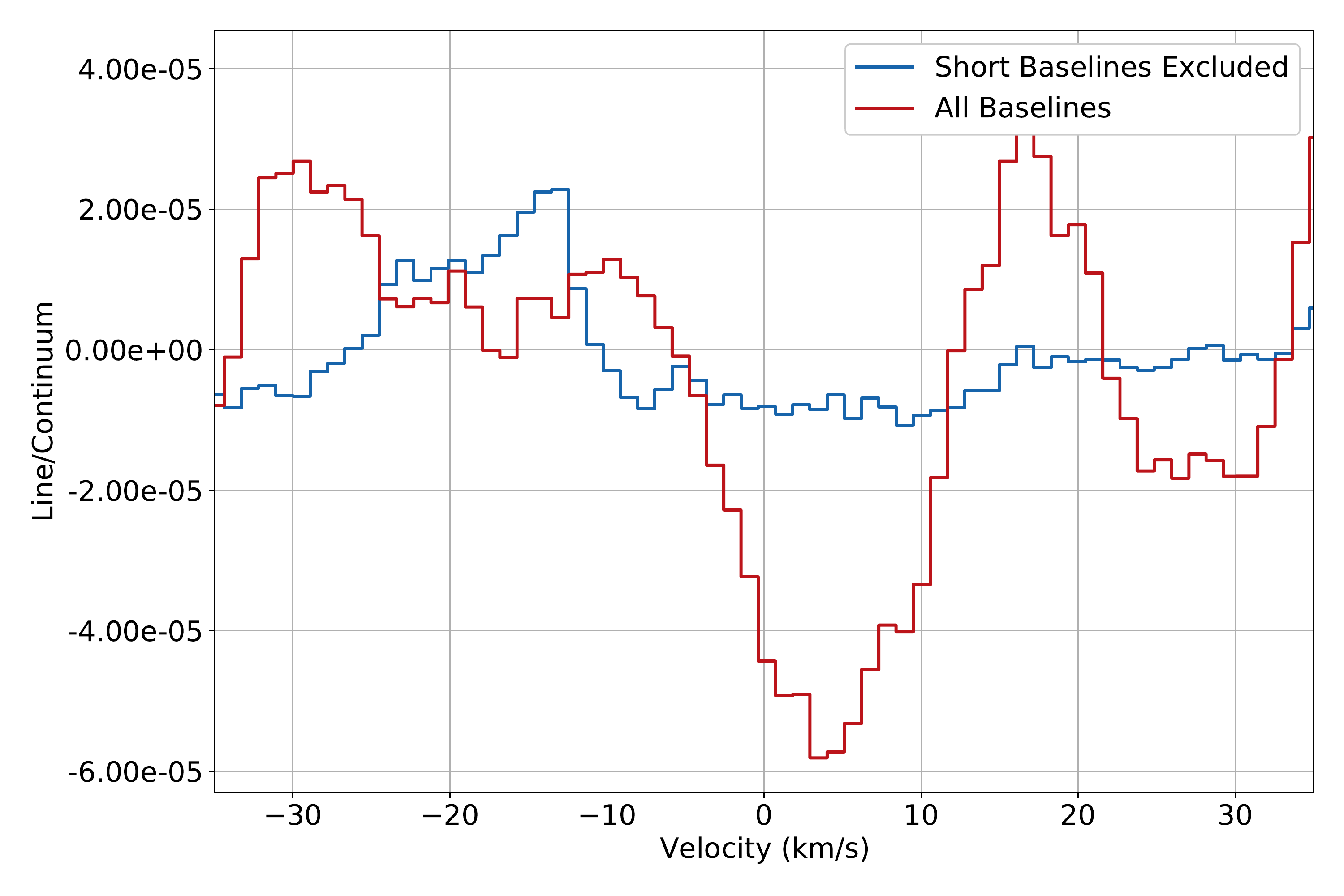} \hfill
    \caption{Narrowband (left) and wideband (right) disk-averaged ALMA spectra for AB and LB datasets both before (top) and after (bottom) revised calibration are shown at a common spectral resolution. All spectra are  continuum-subtracted using 1st order polynomials, and 12th order polynomials are shown fit to the narrowband spectra, as per \citet{Greaves2020}.}
    \label{fig:narrowband}
\end{figure*}

Figure \ref{fig:narrowband} shows narrowband and wideband spectra obtained from the ALMA data of \citet{Greaves2020} binned to a common spectral resolution. The spectra shown in the top row of Figure \ref{fig:narrowband} result from the application of the initial calibration procedures, and those shown in the bottom row results from the use of the revised calibration procedures. Spectra are shown where all ALMA antenna baselines are included and where baselines shorter than 33 meters in length are excluded, following \citet{Greaves2020}. For all spectra, only a first order polynomial is subtracted from the dataset, and a best-fit 12th order polynomial with a 5 km/s exclusion window is superimposed on the narrowband spectra. While a distinct spectral feature would be apparent following the subtraction of the 12th order polynomial from the narrowband spectra obtained with the initial calibration, no such feature is apparent upon correction of the calibration procedures.

\begin{figure*}[tbh!]
    \centering
    \includegraphics[width=0.45\textwidth]{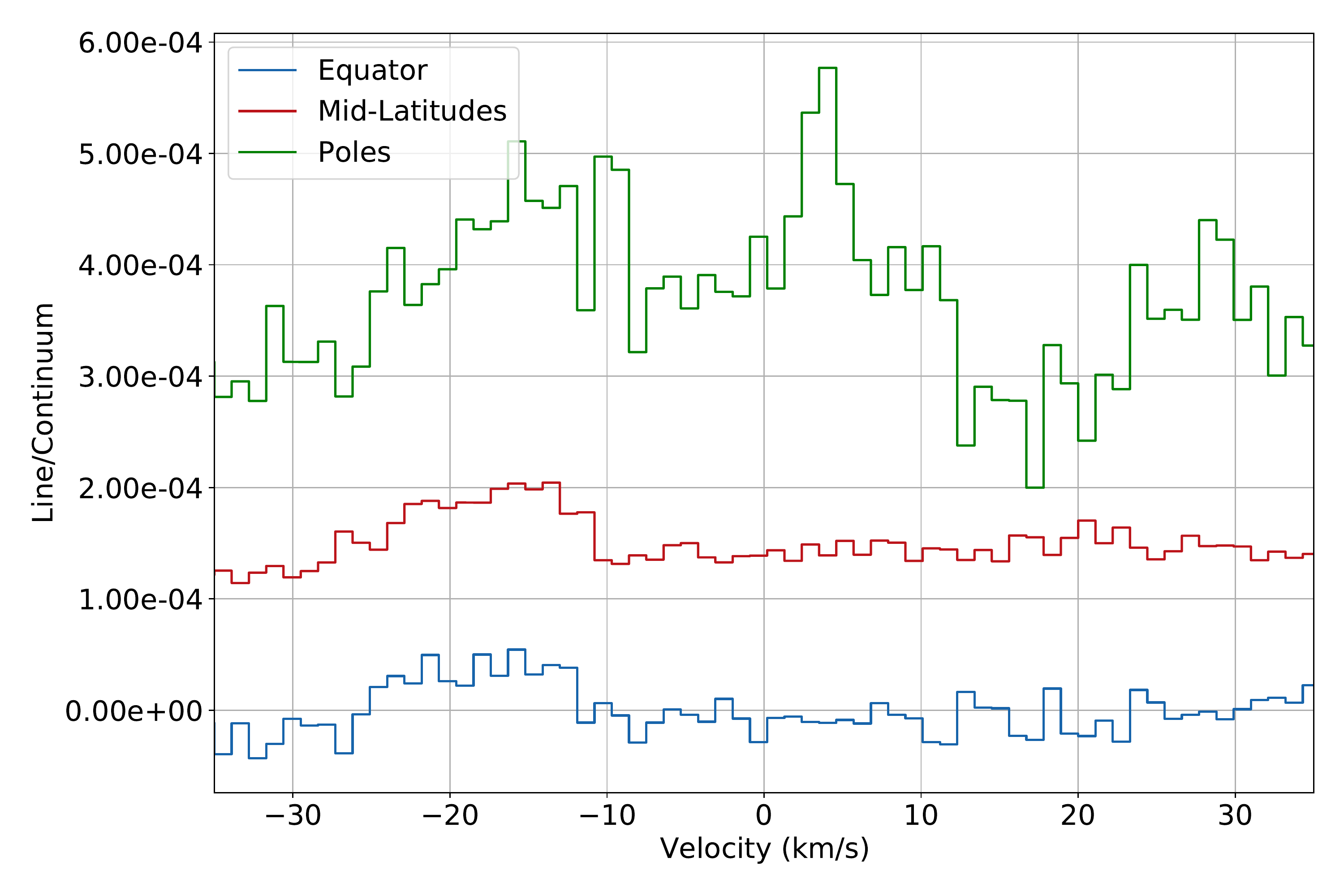}
    \includegraphics[width=0.45\textwidth]{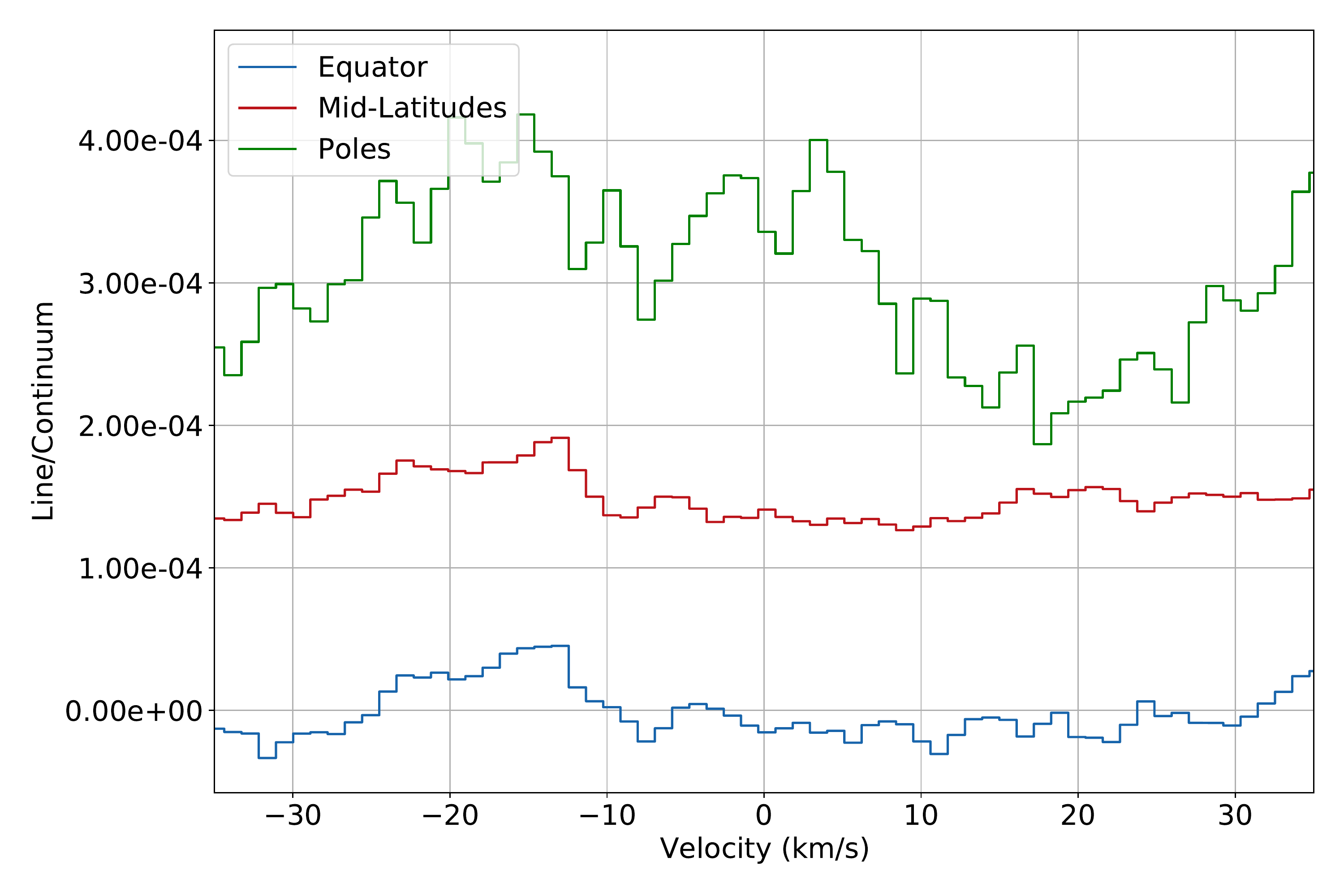} \hfill
    \caption{Narrowband (left) and wideband (right) ALMA spectrum for LB datasets after revised calibration. The spectrum are determined from averaging over the equatorial, mid-latitude, and polar regions (as described in \citet{Greaves2020} Table 1) and they are offset vertically (comparable to \citet{Greaves2020} Fig. 2 ad Extended Data Fig. 2).}
    \label{fig:latdist}
\end{figure*}

Figure \ref{fig:latdist} shows the ALMA spectra determined with revised calibration averaged over the equatorial, mid-latitude, and polar regions. As per \citet{Greaves2020} Table 1, both north and south mid-latitude and polar regions are averaged to form their respective spectra, and the regional spectra are vertically offset for clarity. The narrowband LB spectra in Figure \ref{fig:latdist} are directly comparable to \citet{Greaves2020} Fig. 2 and Extended Data Fig. 2. 

\subsection{Baseline-Dependent Effects and Line Dilution}
\begin{figure*}[tbh!]
    \centering
    \includegraphics[width=0.45\textwidth]{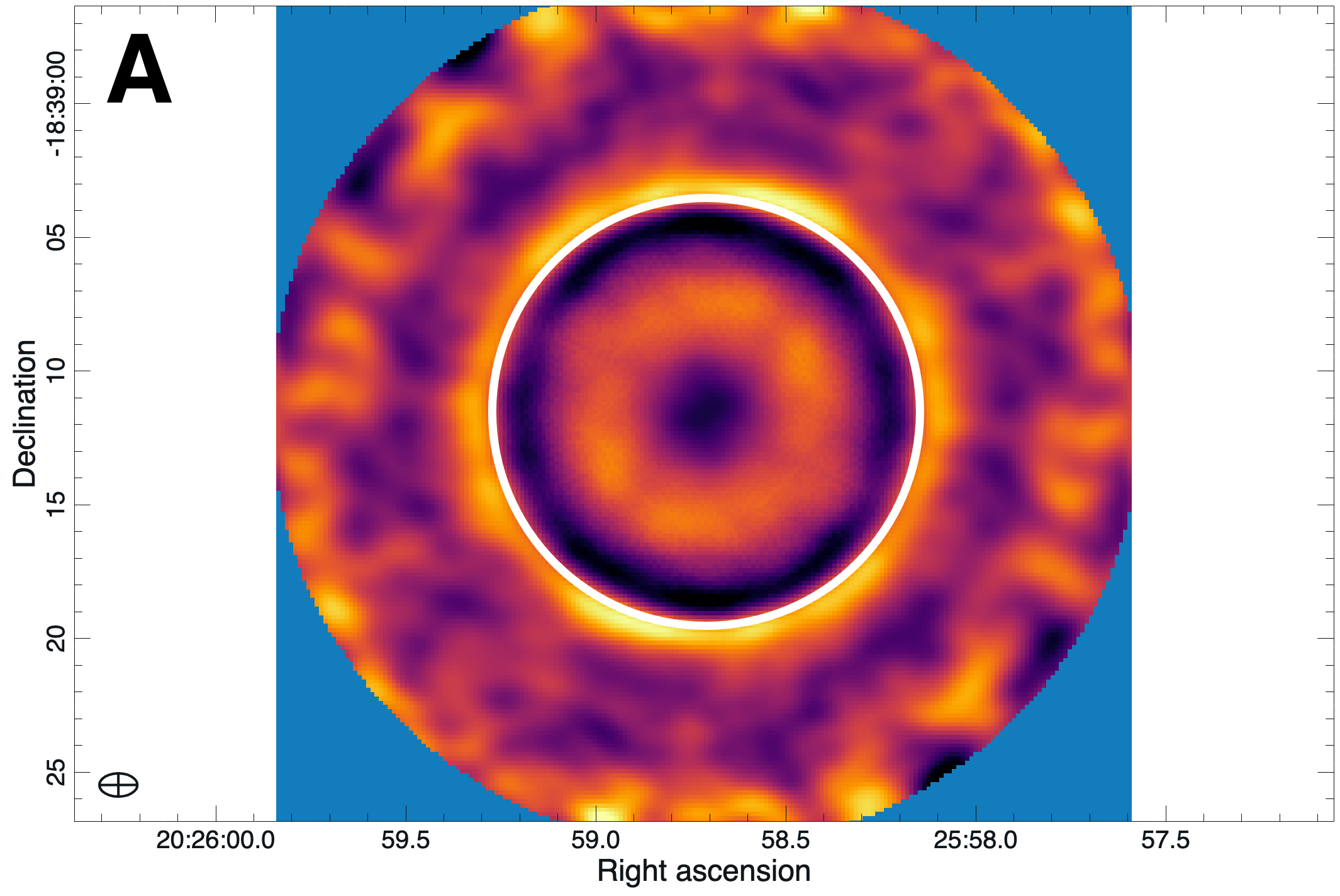}
    \includegraphics[width=0.45\textwidth]{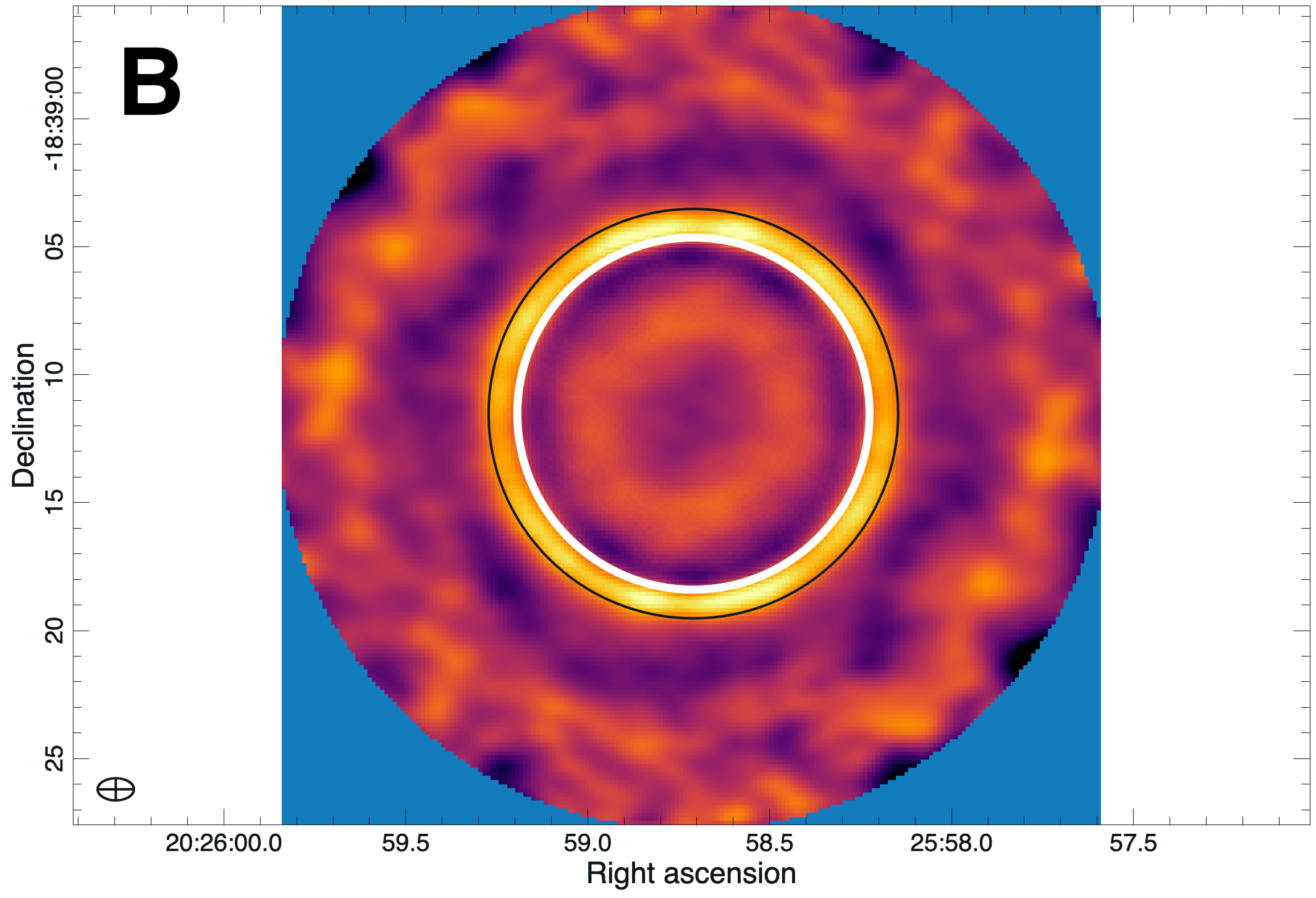}
    \hfill
    \includegraphics[width=0.45\textwidth]{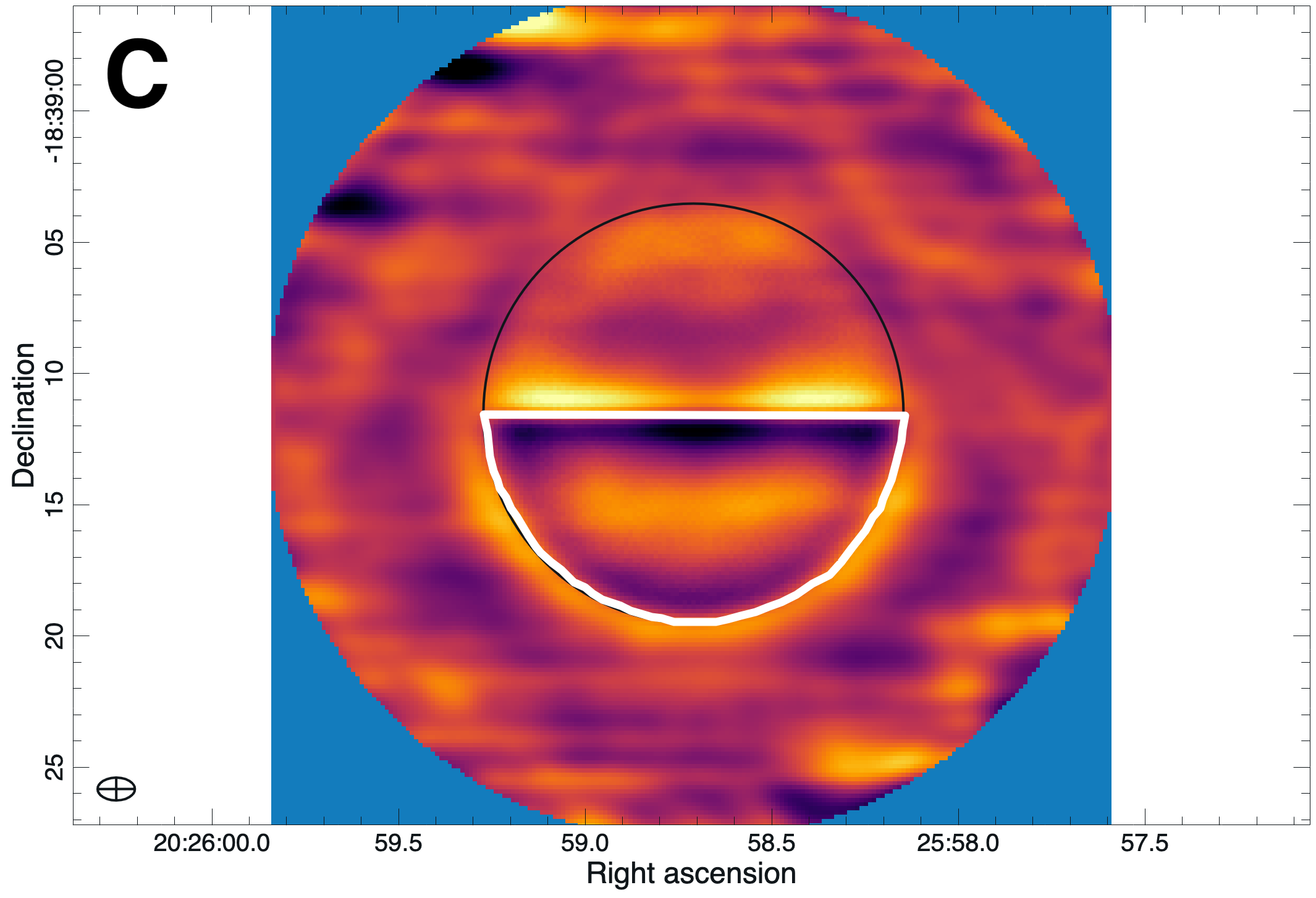}
    \includegraphics[width=0.45\textwidth]{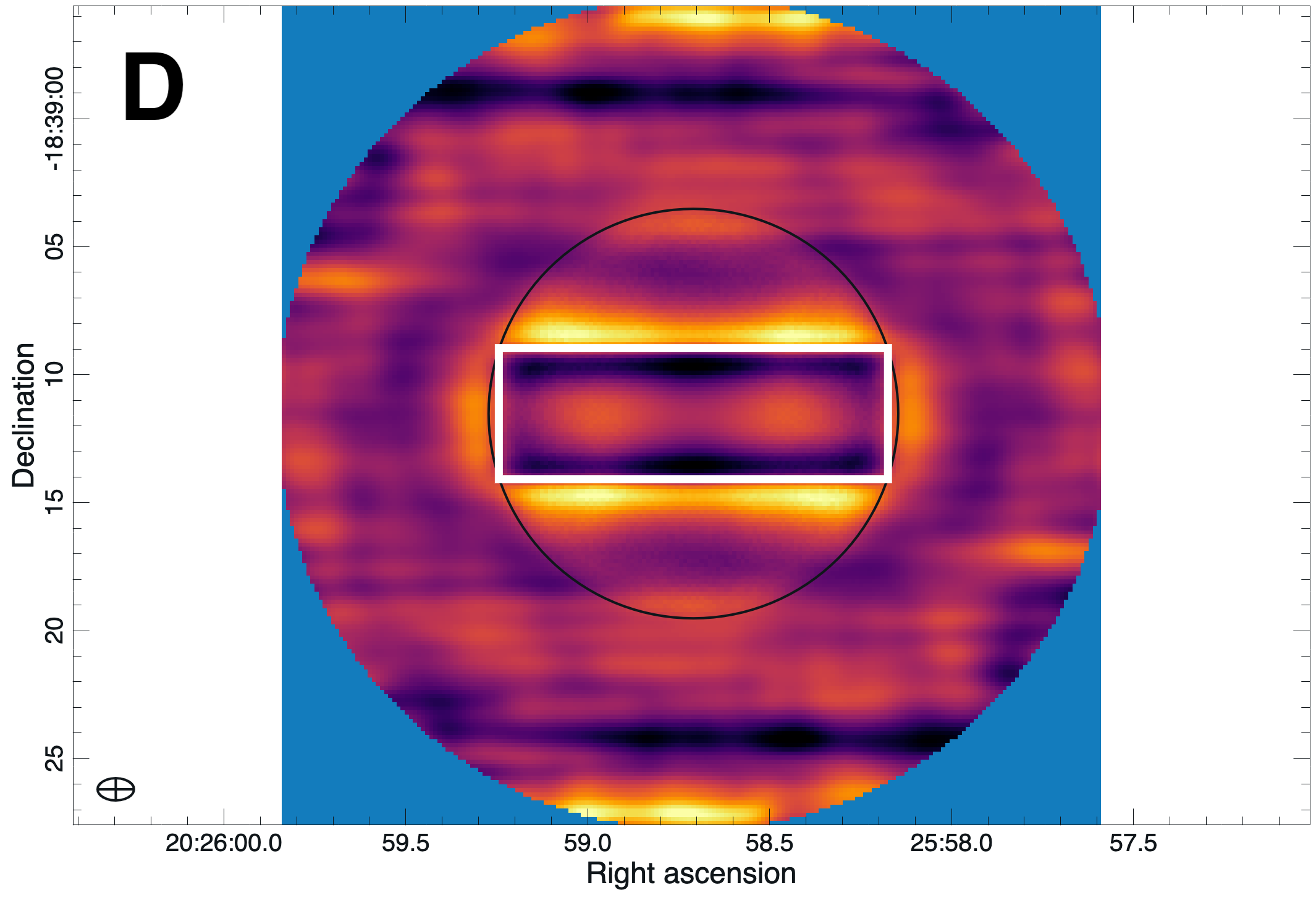}
    \hfill
    \includegraphics[width=0.45\textwidth]{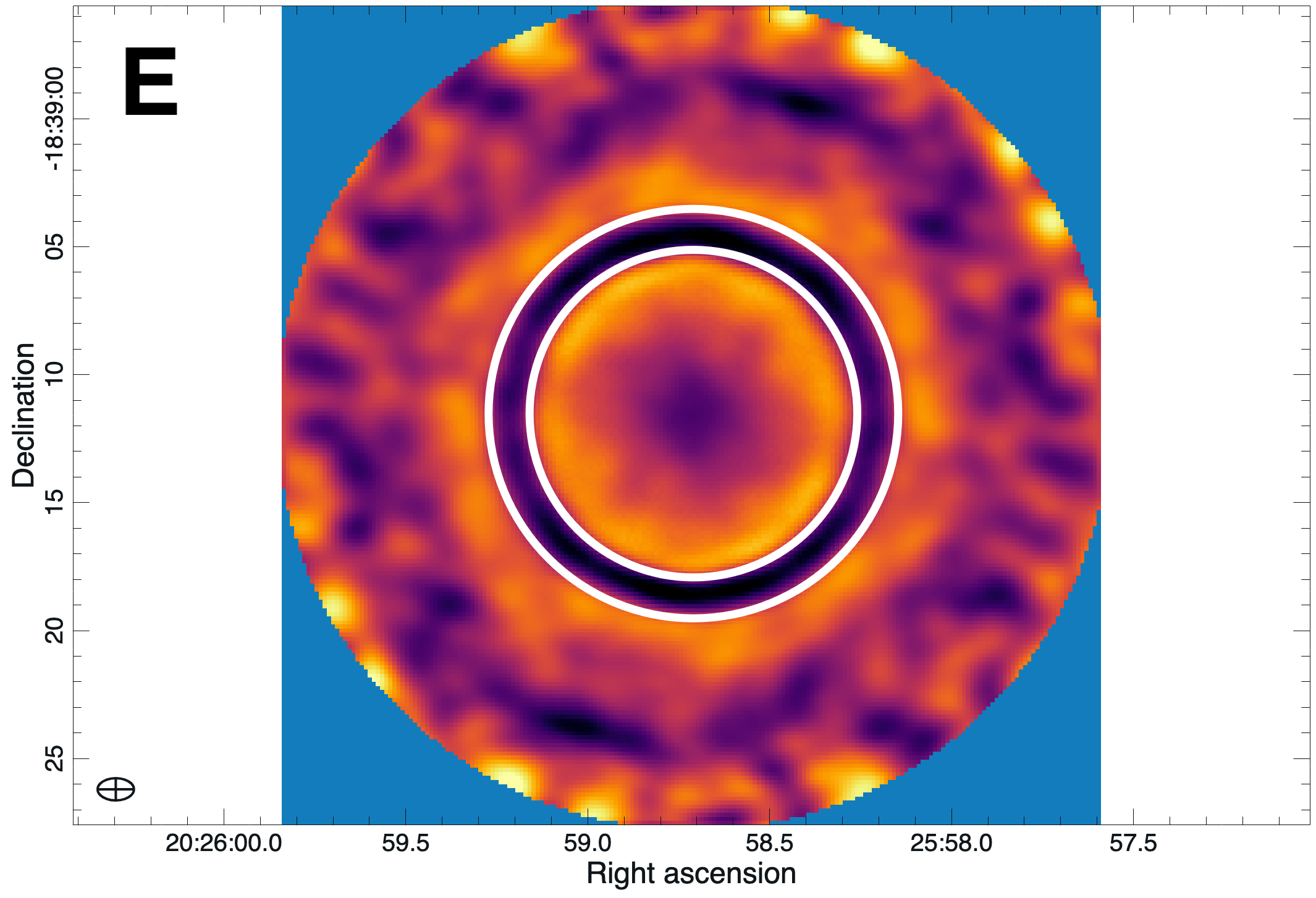}
    \includegraphics[width=0.45\textwidth]{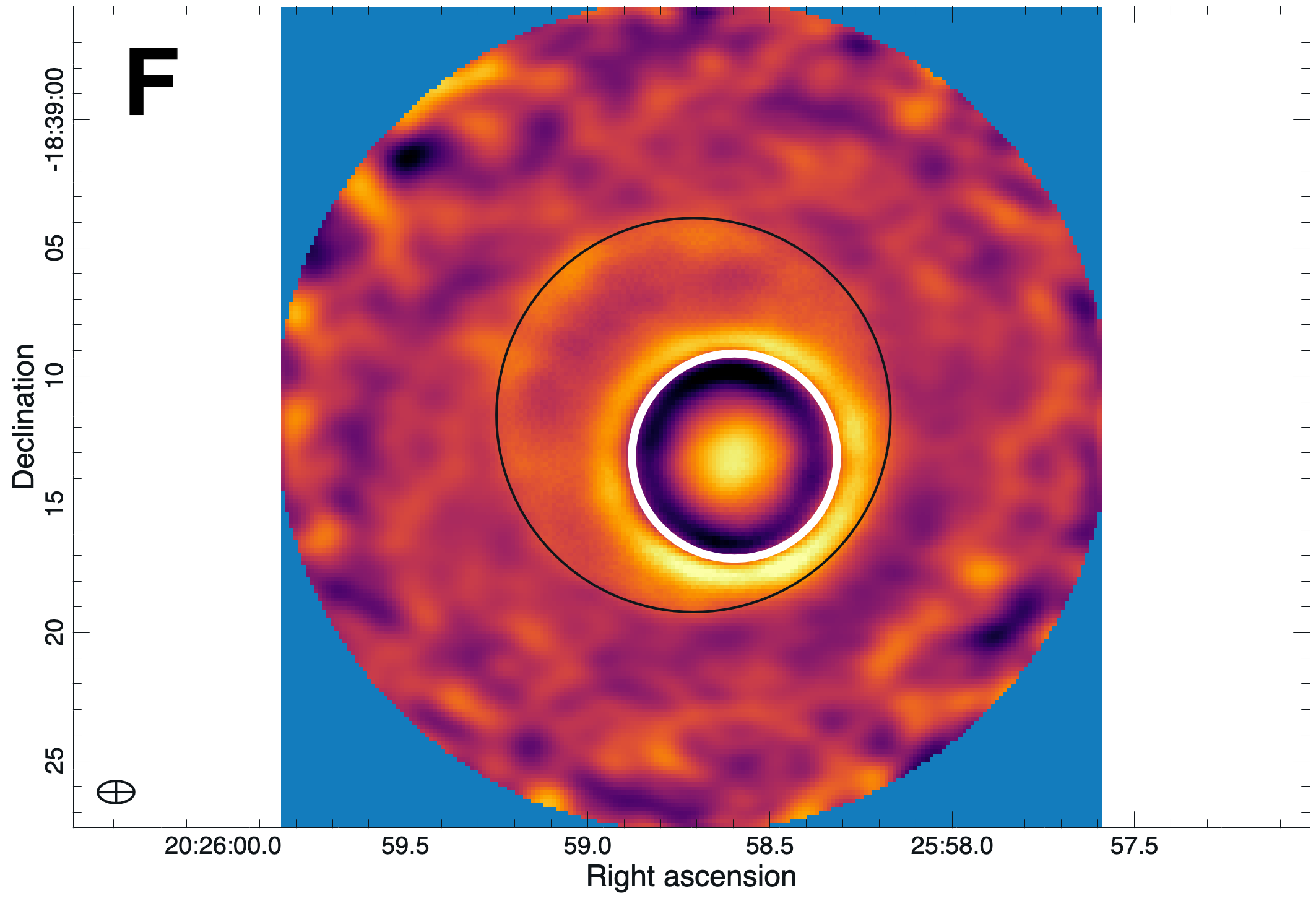}
    \hfill
    \includegraphics[width=0.45\textwidth]{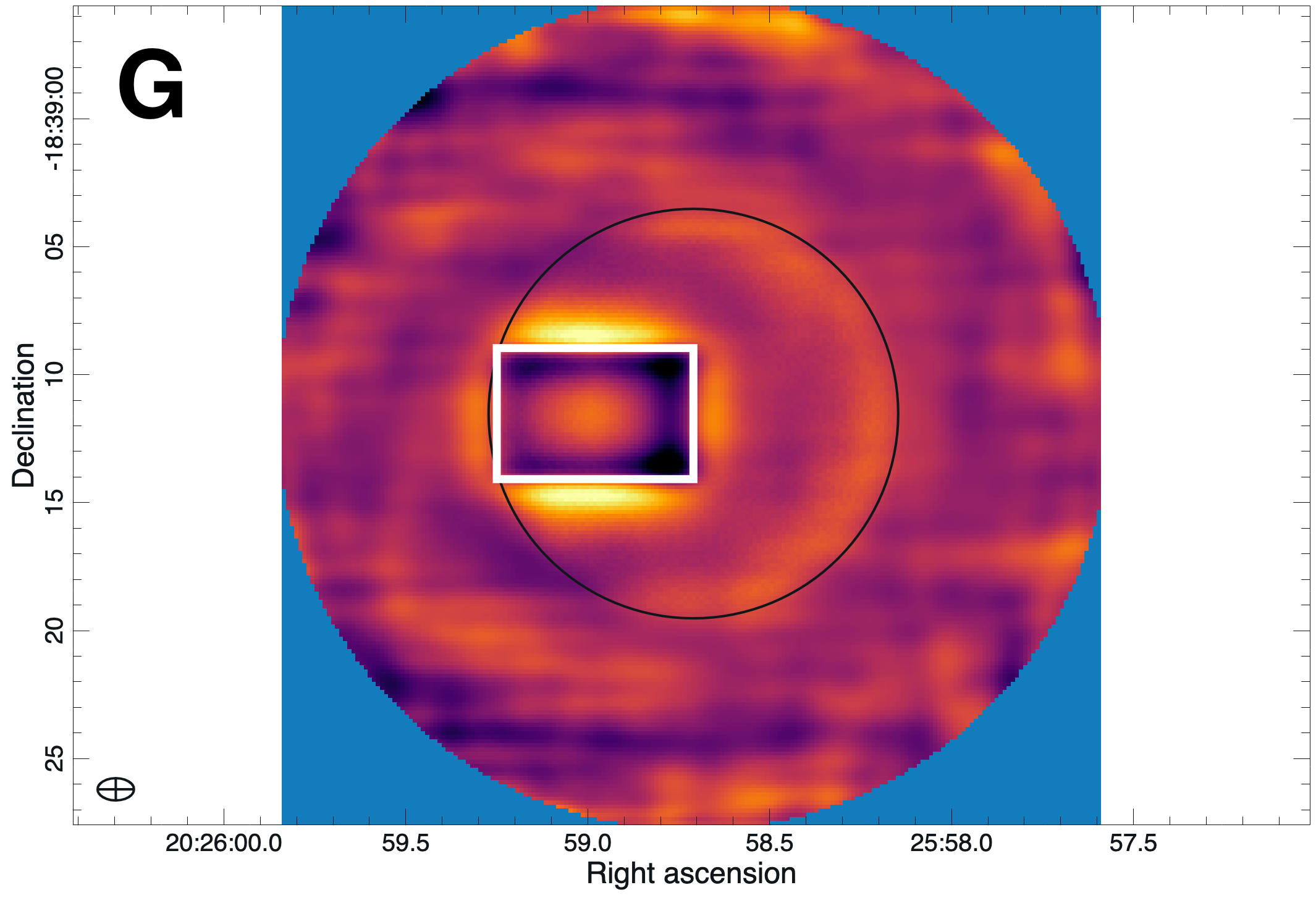}
    \includegraphics[width=0.45\textwidth]{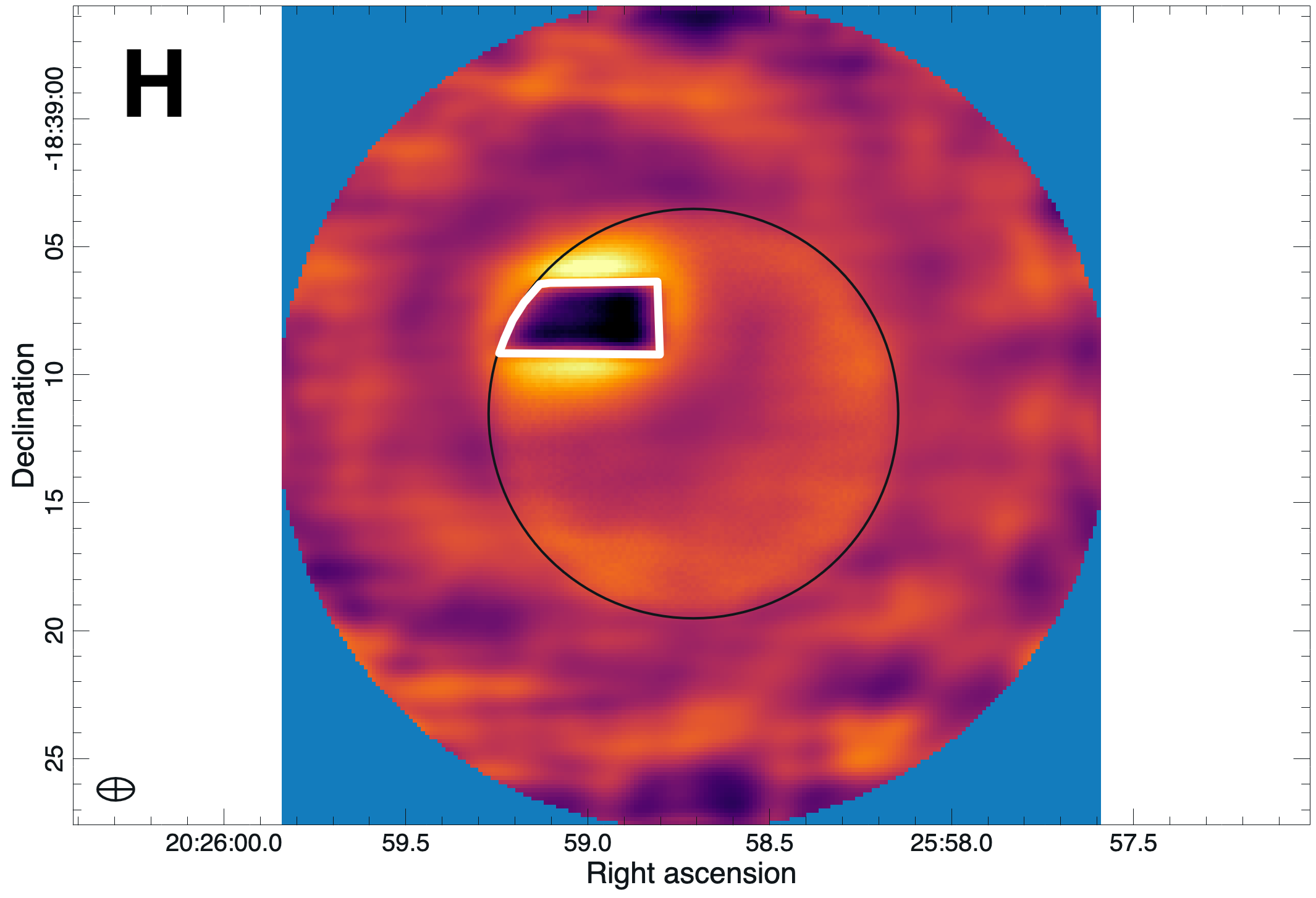}
    \hfill
    \caption{Flux variations across the continuum-subtracted Venus disk at line center for SO$_2$ enhanced models re-imaged using the LB configuration. The extent of Venus is outlined in black, and the spatial regions including SO$_2$ in the starting models are enclosed in white. The color scales of the images are arbitrary; brighter regions correspond to observed emission features, and darker regions correspond to observed absorption features}
    \label{fig:brightnessmap}
\end{figure*}

The dependence of disk-averaged spectral line dilution on the spatial scale of spectral features was studied by re-imaging simulated Venus models enhanced with the SO$_2$ spectra from \citet{Lincowski2020} Figure 2D over a variety of spatial scales and distributions. The apparent dilution of the 267.54 GHz line is discussed here, although there is no change in line dilution over the range of frequencies relevant to the ALMA observations.  The resulting brightness distributions of the re-imaged continuum-subtracted models are illustrated in Figure \ref{fig:brightnessmap}. These model images were re-imaged using the LB dataset baseline coverage and are shown at line center. The regions where SO$_2$ is added to the the starting models are shown in white. Depending on the spatial scale of the spectral features, spatial ripples can be observed across the disk. 

\begin{figure*}[tbh!]
    \centering
    \includegraphics[width=0.45\textwidth]{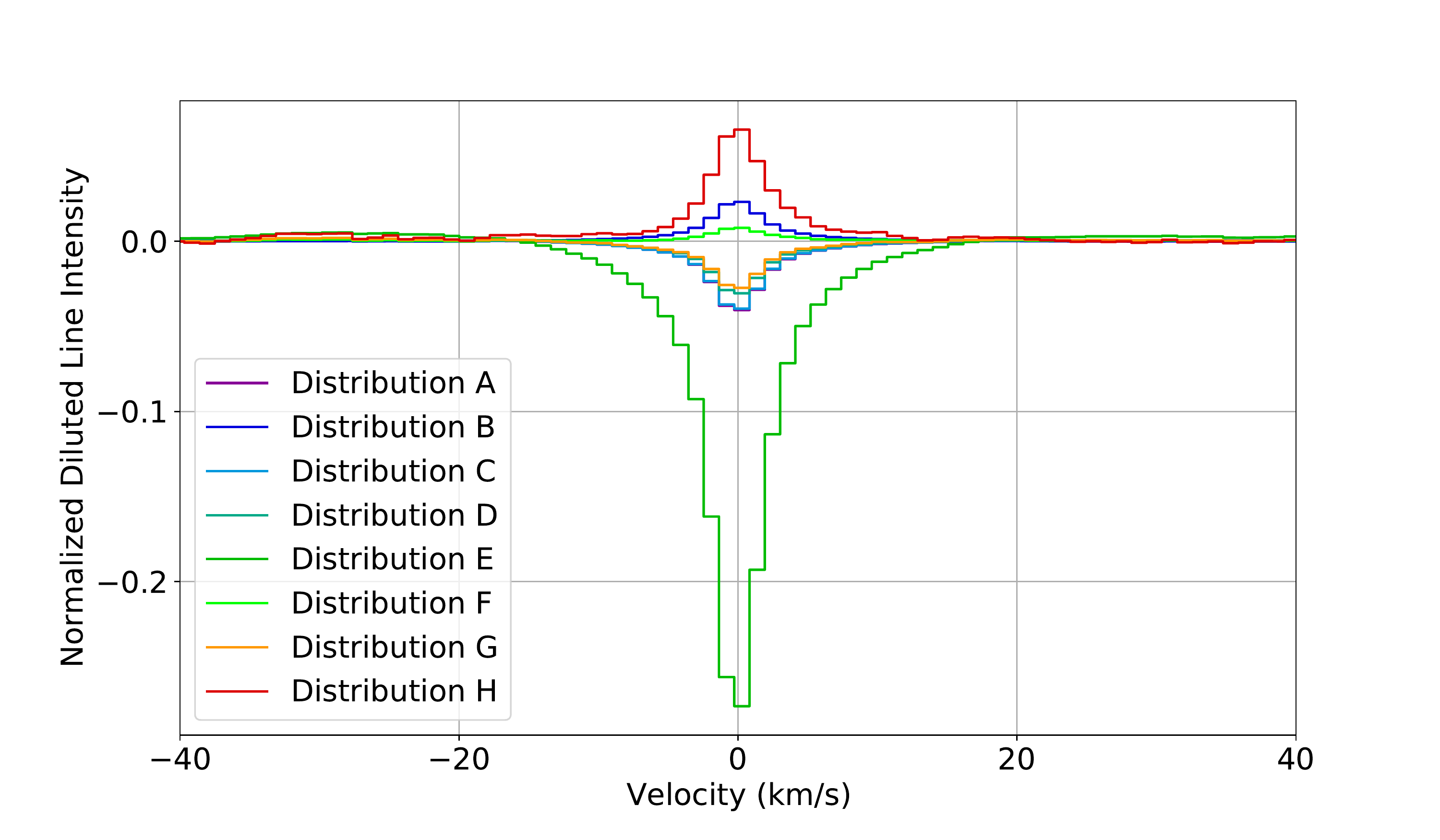}
    \includegraphics[width=0.45\textwidth]{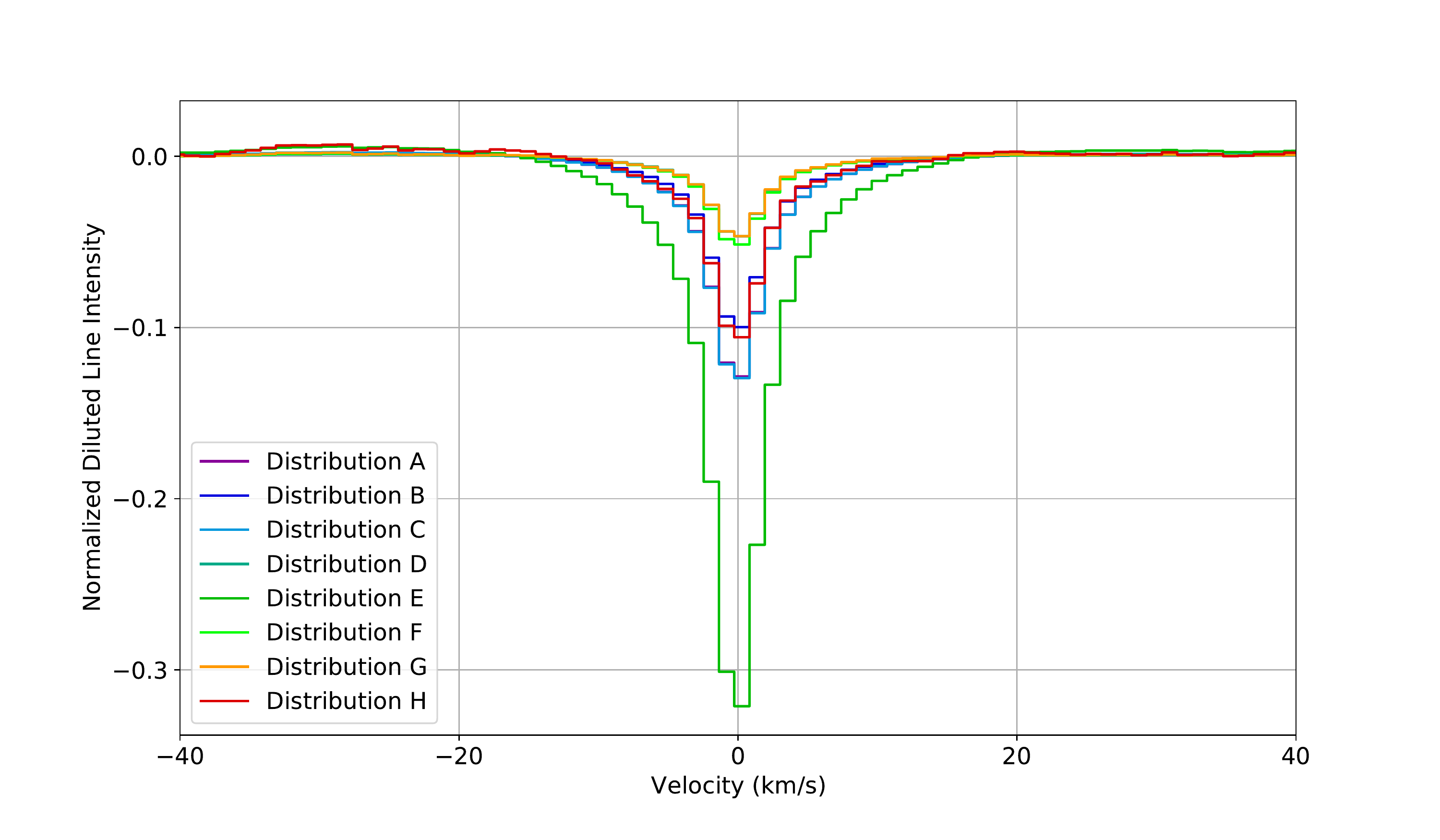}
    \caption{Re-imaged diluted spectral lines normalized by the starting model line intensity for LB (left) and AB (right) datasets. For some cases, the disk-averaged line appears to be in emission. The least diluted configuration corresponds to Distribution E, where the SO$_2$ model spectrum was added to the limb}
    \label{fig:dilution}
\end{figure*}

The apparent disk-averaged line dilutions for both LB and AB datasets are shown in Figure \ref{fig:dilution}. The resulting line intensities are normalized by the disk-averaged intensity of their respective starting models. For both cases, the model SO$_2$ spectral lines are diluted considerably, many by over 90\%. The images excluding short antenna baselines suffer greater line dilution, on the order of 95\% for larger scale distributions. For some cases, the disk-averaged spectral lines appear in emission instead of in absorption due to the ripples resulting from the limited short baseline coverage.

\section{Discussion}
To justify their detection of PH$_3$ in the Venus atmosphere, \citet{Greaves2020} use the following arguments, listed in the order of their subsequent discussion. First, they report detection of a spectral feature at 266.94 GHz from Venus observations using JCMT. Second, they report a repeated detection of this feature from narrowband ALMA observations of Venus. Third, they report simultaneous detection of the 266.94 GHz spectral feature in their ALMA wideband data. Finally, they rule out the possibility that the spectral feature observed can be associated with SO$_2$ by estimating its abundance using a stronger SO$_2$ transition within their wideband ALMA spectrum at 267.54 GHz. Given the results of our independent data reduction and line dilution modeling results, we can address each of these arguments. \par 
First, \citet{Greaves2020} report the detection of a spectral feature at 266.94 GHz from JCMT observations. The authors implicitly suggest that this feature is associated with PH$_3$ and not SO$_2$ due to the non-detection of SO$_2$ lines in the wideband ALMA data observed 2 years later. We find that differences in observation methods between ALMA and JCMT complicate this argument. Since ALMA is a radio interferometer, the maximum recoverable spatial scale and the spatial resolution are determined via the spacing between correlated antenna pairs. Due to the configuration of ALMA at the time, the observed spectrum of gases distributed globally with hemispherical variations, like SO$_2$, will suffer from significant line intensity dilution on the order of 95\%. This problem is compounded by the exclusion of baselines shorter than 33m from the analysis of \citet{Greaves2020}, as shorter baselines are required to measure flux over larger spatial scales. Unlike ALMA, JCMT is a single dish radio telescope measuring the total flux emitted by sources in the direction of antenna pointing independently of the spatial scale (as long as the emission originates from a region observed by the antenna beam). Thus, the JCMT observations do not suffer from scale-dependent spectral line dilution, and so the inferred abundance of SO$_2$ from the \citet{Greaves2020} ALMA wideband data is not applicable here. Recent analyses by \citet{Lincowski2020} and \citet{Villanueva2020} find that nominal abundances of SO$_2$ in the Venus mesosphere are capable of producing the spectral feature at 266.94 GHz as detected by JCMT. In their response paper, \citet{Greaves2020a} argue that the 266.94 GHz feature measured by JCMT could not be SO$_2$ due to the required abundance of 150 ppb, which is considerably higher than millimeter-wavelength measurements \citet{Sandor2010} from 2004-2008 (which are sensitive to emission from higher altitudes). The vertical abundance of SO$_2$, however, is possibly variable up to multiple 100's ppb over longer timescales \citep{Vandaele2017a}. SO$_2$ observations with TEXES/IRTF within a month of the \citet{Greaves2020} observations also reported cloud top abundances of 100's of ppb \citep{Encrenaz2019}.
\par 
Second, \citet{Greaves2020} report the detection of the spectral feature at 266.94 GHz in their ALMA narrowband data. As shown in our Figure \ref{fig:narrowband}, use of the initial, erroneous calibration procedures and subtraction of a 12th order polynomial (with a 5 km/s exclusion window) from the spectral baseline would produce a distinct spectral feature at 266.94 GHz. Following the use of the revised calibration procedures, this spectral feature disappears. The reanalysis of the ALMA data with CASA and AIPS by \citet{Villanueva2020}, with an intermediate correction to the bandpass calibration, results in a non-detection of any spectral feature at 266.94 GHz for the full disk-averaged spectrum. Our reanalysis, using the most recently revised calibration procedures (including a 3rd order polynomial fit for the amplitude bandpass correction) and the same antenna baseline configuration (excluding baselines shorter than 33m), shows no evidence of a spectral feature. This contrasts with \citet{Greaves2020a} who claim that a spectral feature remains following calibration corrections. \citet{Greaves2020a} achieve a lower noise level than our spectra due to their employment of an experimental 12th order polynomial fit for amplitude bandpass calibration. Correcting the bandpass using a higher order polynomial may prove to be questionable in a similar manner to the original 12th order polynomial fit to the spectral baseline described in \citet{Greaves2020} (and criticized by \citet{Snellen2020, Villanueva2020}). All of this considered, the detection of the spectral signal at 266.94 GHz in the narrowband by \citet{Greaves2020} appears to be highly questionable. \par
Third, \citet{Greaves2020} report simultaneous detection of the 266.94 GHz spectral feature in their wideband ALMA data. Their wideband spectrum near 266.94 GHz, however, shows considerable spectral ripples, including a ripple of nearly the same peak magnitude of the detected line at roughly -20 km/s. Unlike \citet{Greaves2020}, we are unable to detect this spectral feature in the ALMA wideband data using either calibration procedure. Furthermore, the spectral ripple in our Figure \ref{fig:narrowband} exhibits a lower peak-to-peak spectral ripple than \citet{Greaves2020} Extended Data Figure 6. This lack of detection from simultaneous observations at comparable noise performance weakens the argument that the narrowband signal is not due to spectral ripple. \par  
Finally, \citet{Greaves2020} suggest that the observed spectral feature cannot be SO$_2$ due to non-detection of a stronger line at 267.54 GHz in the wideband data. Considering their observed spectral ripple, \citet{Greaves2020} assign an upper limit of 10 ppb SO$_2$. This is lower than expected for the Venus cloud top regions, and observations taken only a month later suggest an SO$_2$ abundance of 100s ppb \cite{Encrenaz2019}. From our simulations of spectral line dilution as shown in Figure \ref{fig:dilution}, gases on global and hemispherically symmetric spatial scales can be diluted on the order of 90-95\% due to the ALMA antenna configuration and the subtraction of antenna baselines shorter than 33 m. Thus, SO$_2$ abundances on the order of 100s of ppb would exhibit spectral features below the noise threshold of the ALMA spectra. The successful observation of HDO in the \citet{Greaves2020} narrowband ALMA data, however, does contradict with this assessment, as water is expected to be distributed at similar spatial scales. In their reanalysis, \citet{Greaves2020a} derive a tentative H$_2$O abundance of 23 ppb, which is on the order of 90\% less than prior ALMA observations \citep{Encrenaz2015}. Assuming the \citet{Encrenaz2015} H$_2$O mesospheric abundance of 2.5 ppm, the \citet{Greaves2020a} result is consistent with our line dilution simulations and suggests large scale spatial distribution for H$_2$O.  Future observations of PH$_3$ with ALMA should also include narrowband observations of the 267.54 GHz SO$_2$ spectral line to obtain a more accurate constraint on SO$_2$ abundance.   \par 
Following reassessment of the arguments of \citet{Greaves2020}, it appears unlikely that PH$_3$ was detected in the Venus atmosphere. The non-detection of PH$_3$ at Venus using archival IR observations as reported by \citet{Encrenaz2020} provides further weight to this argument. While \citet{Greaves2020a} detect a weak feature at 266.94 GHz in their reanalysis, they admit that the strength of this feature would be consistent with SO$_2$ contamination. A stronger answer to this question can be found through continued ground and space-based observation of Venus at millimeter, sub-millimeter, and infrared wavelengths near PH$_3$ spectral features. These observations should also target features of possible spectral contaminants.

\section{Conclusions}

We have reanalyzed the wideband and narrowband ALMA data of \citet{Greaves2020} to investigate the robustness of their detection of PH$_3$. We have also simulated the spectral line dilution resulting from their ALMA configuration for several different distributions of an absorbing gas at varying spatial scales. We find that a spectral feature near 266.94 GHz in the narrowband data becomes apparent using the initial calibration procedures of \citet{Greaves2020} and subtraction of a 12th order polynomial. Upon application of the revised calibration procedures, however, no such spectral feature is detected. We are also unable to detect a spectral feature in the wideband ALMA spectra. Additionally, we have used the line dilution for gases distributed at a global scale to find that the non-detection of SO$_2$ in the wideband ALMA data is potentially consistent with a nominal cloud top abundance far greater than the 10 ppb SO$_2$ upper limit suggested by \citet{Greaves2020} (see also the companion study in \citet{Lincowski2020}). 
These findings, along with the recent papers by \citet{Encrenaz2020}, \citet{Snellen2020}, \citet{Lincowski2020}, and \citet{Villanueva2020} undermine the reported detection of PH$_3$ by \citet{Greaves2020, Greaves2020a} and its possible biogenic origin. It is clear that further studies of Venus at millimeter, sub-millimeter, and infrared wavelengths with ground or space-based observatories are necessary to further investigate the possibility of PH$_3$ at Venus. 

\section{Acknowledgements}
We thank the authors of the original study for publicly sharing their initial calibration and imaging scripts. We also thank ESO ALMA Regional Center for their efforts during QA3 assessment of the original dataset and for making the updated calibration procedures available via the ALMA archive. Venus was observed under ALMA Director’s Discretionary Time program 2018.A.0023.S. ALMA is a partnership of ESO (representing its member states), NSF (USA) and NINS (Japan), together with NRC (Canada), MOST and ASIAA (Taiwan), and KASI (Republic of Korea), in cooperation with the Republic of Chile. The JAO is operated by ESO, AUI/NRAO and NAOJ. We thank J. Pearson for helpful comments on the manuscript. Part of this work was conducted at the Jet Propulsion Laboratory, California Institute of Technology, under contract with NASA. Part of this work was performed by the Virtual Planetary Laboratory Team, which is a member of the NASA Nexus for Exoplanet System Science, and funded via NASA Astrobiology Program Grant 80NSSC18K0829. 

\software{CARTA \citep{Comrie2020}, CASA \citep{McMullin2007}, Matplotlib \citep{Hunter2007}, Numpy \citep{Harris2020}}

\bibliographystyle{aasjournal} \bibliography{library.bib}

\begin{thebibliography}{}
\expandafter\ifx\csname natexlab\endcsname\relax\def\natexlab#1{#1}\fi
\providecommand{\url}[1]{\href{#1}{#1}}
\providecommand{\dodoi}[1]{doi:~\href{http://doi.org/#1}{\nolinkurl{#1}}}
\providecommand{\doeprint}[1]{\href{http://ascl.net/#1}{\nolinkurl{http://ascl.net/#1}}}
\providecommand{\doarXiv}[1]{\href{https://arxiv.org/abs/#1}{\nolinkurl{https://arxiv.org/abs/#1}}}

\bibitem[{Bains {et~al.}(2020)Bains, Petkowski, Seager, Ranjan, Sousa-Silva,
  Rimmer, Zhan, Greaves, \& Richards}]{Bains2020}
Bains, W., Petkowski, J.~J., Seager, S., {et~al.} 2020, Astrobiology.
\newblock \url{http://arxiv.org/abs/2009.06499}

\bibitem[{Butler \& Bastian(1999)}]{Butler1999}
Butler, B.~J., \& Bastian, T.~S. 1999, Solar System Objects

\bibitem[{Comrie {et~al.}(2020)Comrie, Wang, Hsu, Moraghan, Harris, Pang,
  Pińska, Chiang, Simmonds, Chang, Jan, \& Lin}]{Comrie2020}
Comrie, A., Wang, K.-S., Hsu, S.-C., {et~al.} 2020, CARTA: The Cube Analysis
  and Rendering Tool for Astronomy,  Zenodo, \dodoi{10.5281/zenodo.4034416}

\bibitem[{Encrenaz {et~al.}(2015)Encrenaz, Moreno, Moullet, Lellouch, \&
  Fouchet}]{Encrenaz2015}
Encrenaz, T., Moreno, R., Moullet, A., Lellouch, E., \& Fouchet, T. 2015,
  Planetary and Space Science, 113-114, 275, \dodoi{10.1016/j.pss.2015.01.011}

\bibitem[{Encrenaz {et~al.}(2019)Encrenaz, Greathouse, Marcq, Sagawa, Widemann,
  Bézard, Fouchet, Lefèvre, Lebonnois, Atreya, Lee, Giles, \&
  Watanabe}]{Encrenaz2019}
Encrenaz, T., Greathouse, T.~K., Marcq, E., {et~al.} 2019, Astronomy \&
  Astrophysics, 623, A70, \dodoi{10.1051/0004-6361/201833511}

\bibitem[{Encrenaz {et~al.}(2020)Encrenaz, Greathouse, Marcq, Widemann,
  Bézard, Fouchet, Giles, Sagawa, Greaves, \& Sousa-Silva}]{Encrenaz2020}
---. 2020, 1.
\newblock \url{http://arxiv.org/abs/2010.07817}

\bibitem[{Greaves {et~al.}(2020{\natexlab{a}})Greaves, Richards, Bains, Rimmer,
  Sagawa, Clements, Seager, Petkowski, Sousa-Silva, Ranjan, Drabek-Maunder,
  Fraser, Cartwright, Mueller-Wodarg, Zhan, Friberg, Coulson, Lee, \&
  Hoge}]{Greaves2020}
Greaves, J.~S., Richards, A. M.~S., Bains, W., {et~al.} 2020{\natexlab{a}},
  Nature Astronomy, \dodoi{10.1038/s41550-020-1174-4}

\bibitem[{Greaves {et~al.}(2020{\natexlab{b}})Greaves, Richards, Bains, Rimmer,
  Clements, Seager, Petkowski, Sousa-Silva, Ranjan, \& Fraser}]{Greaves2020a}
---. 2020{\natexlab{b}}, Nature Astronomy, 1.
\newblock \url{http://arxiv.org/abs/2011.08176}

\bibitem[{Harris {et~al.}(2020)Harris, Millman, van~der Walt, Gommers,
  Virtanen, Cournapeau, Wieser, Taylor, Berg, Smith, Kern, Picus, Hoyer, van
  Kerkwijk, Brett, Haldane, del R'io, Wiebe, Peterson, G'erard-Marchant,
  Sheppard, Reddy, Weckesser, Abbasi, Gohlke, \& Oliphant}]{Harris2020}
Harris, C.~R., Millman, K.~J., van~der Walt, S.~J., {et~al.} 2020, Nature, 585,
  357, \dodoi{10.1038/s41586-020-2649-2}

\bibitem[{Hunter(2007)}]{Hunter2007}
Hunter, J.~D. 2007, Computing in Science \& Engineering, 9, 90,
  \dodoi{10.1109/MCSE.2007.55}

\bibitem[{Limaye {et~al.}(2018)Limaye, Mogul, Smith, Ansari, Słowik, \&
  Vaishampayan}]{Limaye2018}
Limaye, S.~S., Mogul, R., Smith, D.~J., {et~al.} 2018, Astrobiology, 18,
  ast.2017.1783, \dodoi{10.1089/ast.2017.1783}

\bibitem[{Lincowski {et~al.}(2020)Lincowski, Meadows, Crisp, Akins,
  Schwieterman, Arney, Wong, Steffes, Parenteau, \&
  Domagal-Goldman}]{Lincowski2020}
Lincowski, A.~P., Meadows, V.~S., Crisp, D., {et~al.} 2020, Astrophysical
  Journal Letters

\bibitem[{Marcq {et~al.}(2018)Marcq, Mills, Parkinson, \& Vandaele}]{Marcq2018}
Marcq, E., Mills, F.~P., Parkinson, C.~D., \& Vandaele, A.~C. 2018, Space
  Science Reviews, 214, 10, \dodoi{10.1007/s11214-017-0438-5}

\bibitem[{McMullin {et~al.}(2007)McMullin, Waters, Schiebel, Young, \&
  Golap}]{McMullin2007}
McMullin, J.~P., Waters, B., Schiebel, D., Young, W., \& Golap, K. 2007, in
  CASA Architecture and Applications, Vol. 376, 127--130

\bibitem[{Remijan {et~al.}(2019)Remijan, Biggs, Cortes, Dent, Francesco,
  Fomalont, Hales, Kameno, Mason, Philips, Stoehr, Vilaro, \&
  Villard}]{Remijan2019}
Remijan, A., Biggs, A., Cortes, P.~A., {et~al.} 2019, ALMA Technical Handbook

\bibitem[{Sandor {et~al.}(2010)Sandor, Clancy, Moriarty-Schieven, \&
  Mills}]{Sandor2010}
Sandor, B.~J., Clancy, R.~T., Moriarty-Schieven, G., \& Mills, F.~P. 2010,
  Icarus, 208, 49, \dodoi{10.1016/j.icarus.2010.02.013}

\bibitem[{Seager {et~al.}(2020)Seager, Petkowski, Gao, Bains, Bryan, Ranjan, \&
  Greaves}]{Seager2020}
Seager, S., Petkowski, J.~J., Gao, P., {et~al.} 2020, Astrobiology, 21, 1,
  \dodoi{10.1089/ast.2020.2244}

\bibitem[{Snellen {et~al.}(2020)Snellen, Guzman-Ramirez, Hogerheijde, Hygate,
  \& van~der Tak}]{Snellen2020}
Snellen, I. A.~G., Guzman-Ramirez, L., Hogerheijde, M.~R., Hygate, A. P.~S., \&
  van~der Tak, F. F.~S. 2020, Astronomy and Astrophysics, 1.
\newblock \url{http://arxiv.org/abs/2010.09761}

\bibitem[{Thompson(2020)}]{Thompson2020}
Thompson, M.~A. 2020, Monthly Notices of the Royal Astronomical Society, 4, 1.
\newblock \url{http://arxiv.org/abs/2010.15188}

\bibitem[{Vandaele {et~al.}(2017)Vandaele, Korablev, Belyaev, Chamberlain,
  Evdokimova, Encrenaz, Esposito, Jessup, Lefèvre, Limaye, Mahieux, Marcq,
  Mills, Montmessin, Parkinson, Robert, Roman, Sandor, Stolzenbach, Wilson, \&
  Wilquet}]{Vandaele2017a}
Vandaele, A.~C., Korablev, O., Belyaev, D.~A., {et~al.} 2017, Icarus, 295, 1,
  \dodoi{10.1016/j.icarus.2017.05.001}

\bibitem[{Villanueva {et~al.}(2020)Villanueva, Cordiner, Irwin, de~Pater,
  Butler, M., Milam, Nixon, Luszcz-Cook, Wilson, Kofman, Liuzzi, Faggi,
  Fauchez, Lippi, Cosentino, Thelen, Moullet, Hartogh, Molter, Charnley, Arney,
  Mandell, Biver, Vandaele, de~Kleer, \& Kopparapu}]{Villanueva2020}
Villanueva, G., Cordiner, M., Irwin, P., {et~al.} 2020, Nature Astronomy

\end{thebibliography}

\end{document}